\documentclass[journal, twoside,web]{ieeecolor}
\usepackage{lcsys}
\usepackage{cite}
\usepackage{amsmath,amssymb,amsfonts}
\usepackage{graphicx}
\usepackage{textcomp}
\usepackage{fancyhdr}

\usepackage{graphics,graphicx,psfrag,array,eepic}
\usepackage{multirow}
\usepackage{hhline}
\usepackage{mathtools}
\usepackage[strict]{changepage}
\usepackage{epsfig}
\usepackage{amsmath}
\usepackage{amssymb}
\usepackage{xfrac}
\usepackage{float}
\restylefloat{table}

\usepackage{multicol}
\usepackage{color}
\usepackage{balance}
\usepackage{cite}
\usepackage{booktabs} 
\usepackage{caption,booktabs}
\usepackage[tableposition=top]{caption}
\usepackage{subcaption}

\usepackage{lipsum}
\usepackage{comment}
\usepackage{amssymb}
\usepackage{graphicx}

\usepackage{comment}
\usepackage{bm}

\usepackage{algorithm}
\usepackage[noend]{algpseudocode}

\usepackage{mathtools, cuted}
\usepackage[thinc]{esdiff}

\usepackage{psfrag}
\usepackage{array}
\usepackage{latexsym,theorem}
\usepackage{amsfonts,amssymb,amsbsy}
\usepackage{tikz}
\usetikzlibrary{decorations.shapes,shapes.geometric,shadows,arrows,automata,positioning,calendar,mindmap,backgrounds,scopes,chains,er,3d,matrix,fit}
\usepackage{tikz-cd}
\usepackage{pgfplots}
\usetikzlibrary{intersections, pgfplots.fillbetween}

\newcommand{\R}{{\rm I\!R}}

\newcommand{\N}{{\rm I\!N}}

\newcommand{\cU}{\mathcal{U}}
\newcommand{\cX}{\mathcal{X}}

\newcommand{\cE}{\mathcal{E}}

\newcommand{\cV}{\mathcal{V}}

\newcommand{\cR}{\mathcal{R}}

\newcommand{\cZ}{\mathcal{Z}}

\newcommand{\cD}{\mathcal{D}}
\newcommand{\cW}{\mathcal{W}}

\newtheorem{lemma}{Lemma}
\newtheorem{definition}{Definition}
\newtheorem{theorem}{Theorem}

\newtheorem{remark}{Remark}

\usepackage[british]{babel}
\usepackage{cite}
\usepackage{amsmath,amssymb,amsfonts}
\usepackage{graphicx}
\usepackage{textcomp}
\usepackage{mathrsfs}
\usepackage{url}
\usepackage{bm}
\usepackage{nicefrac}
\usepackage{hyperref}
\usepackage{pgfplots}
\allowdisplaybreaks
\usepackage{alphalph,etoolbox}
\usepackage{mathtools,subcaption,float}
\usepackage{pgfplots}
\usepgfplotslibrary{colorbrewer}
\pgfplotsset{compat = 1.15} 
\usetikzlibrary{pgfplots.statistics, pgfplots.colorbrewer} 
\usepackage{pgfplotstable}

\newcommand{\prb}[1]{\Pr\{#1\}}
\DeclareMathOperator{\Quantile}{Quantile}
\newcommand{\quan}[2]{\Quantile(\{#1\},#2)}

\def\BibTeX{{\rm B\kern-.05em{\sc i\kern-.025em b}\kern-.08em
    T\kern-.1667em\lower.7ex\hbox{E}\kern-.125emX}}
\fancypagestyle{plain}{ 
	\fancyhf{} 
	\fancyhead[C]{To appear in IEEE Control Systems Letters (L-CSS)} 
}

\begin{document}

\title{Conformal Prediction for Distribution-free Optimal Control of Linear Stochastic Systems}
\author{Eleftherios E. Vlahakis, Lars Lindemann, \IEEEmembership{Member, IEEE}, Pantelis Sopasakis, and \\ Dimos V. Dimarogonas, \IEEEmembership{Fellow, IEEE} 
\thanks{This work was supported by the Swedish
Research Council (VR), the Knut \& Alice Wallenberg Foundation (KAW), the Horizon Europe Grant SymAware and the ERC Consolidator Grant LEAFHOUND.}
\thanks{E.E. Vlahakis and D.V. Dimarogonas are with the Division of Decision and Control Systems, School of Electrical Engineering and Computer Science, KTH Royal Institute of Technology, 10044, Stockholm, Sweden. Email: {\tt\small\{vlahakis,dimos\}@kth.se}. L. Lindemann is with the Thomas Lord Department of Computer Science and the Ming Hsieh Department of Electrical and Computer Engineering, Viterbi School of Engineering, University of Southern California,  Los Angeles, 90089, CA, USA. Email: {\tt\small llindema@usc.edu}. P. Sopasakis is with the School of Electronics, Electrical Engineering and Computer Science, Queen's University Belfast, Northern Ireland, UK. Email: {\tt\small p.sopasakis@qub.ac.uk}
}}

\maketitle
\thispagestyle{plain}

\begin{abstract}
We address an optimal control problem for linear stochastic systems with unknown noise distributions and joint chance constraints using conformal prediction. Our approach involves designing a feedback controller to maintain an error system within a prediction region (PR). We define PRs as sublevel sets of a nonconformity score over error trajectories, enabling the handling of joint chance constraints. We propose two methods to design feedback control and PRs: one through direct optimization over error trajectory samples, and the other indirectly using the $S$-procedure with a disturbance ellipsoid obtained from data. By tightening constraints with PRs, we solve a relaxed problem to synthesize a feedback policy. Our method ensures reliable probabilistic guarantees based on marginal coverage, independent of data size. 
\end{abstract}


\section{Introduction}
\label{sec:introduction}

\IEEEPARstart{P}{robabilistic} guarantees play a crucial role in many applications that involve stochastic disturbances and safety. As chance-constrained problems are generally non-convex and computationally intractable, most approaches solve deterministic relaxations by applying constraint-tightening techniques based on available information about uncertainty \cite{Mesbah2016}. Constraint tightening for probabilistic satisfaction has been studied when the underlying probability distribution is known \cite{Kouvaritakis2010explicit}, in cases of bounded disturbances \cite{CannonTAC2011}, or using probabilistic reachable sets \cite{HewingCDC2018, KohlerLCSS2022, VlahakisCDC24}. These approaches may be restricted to Gaussian settings, involve computationally costly operations, or often rely on the multivariate Chebyshev inequality, which can lead to conservative bounds. In contrast, sampling-based methods such as \cite{Prandini2012, LorenzenTAC2017, HewingLCSS2020} can be significantly more flexible in relaxing chance-constrained problems by utilizing available disturbance samples, termed scenarios, and can provide formal guarantees based on scenario optimization (SO) \cite{Calafiore2005, Campi2008}.

An alternative distribution-free framework for streamlined uncertainty quantification, equipped with formal guarantees, is provided by conformal prediction (CP). Originally introduced by Vovk and Shafer \cite{vovk2005algorithmic, shafer2008tutorial}, CP uses a calibration dataset to infer prediction regions for a test dataset with a specified probability, while remaining distribution-agnostic. CP has been applied in machine learning \cite{angelopoulos2022gentle}, chance-constrained optimization \cite{zhao2024}, and control for safety verification and planning \cite{Muthali2023, Stamouli24, LindemannRAL2023}, see \cite{lindemann2024CPsurvey} for a recent survey.

Leveraging its distribution independence, we apply CP to tackle an optimal control problem for linear stochastic systems under joint chance constraints, assuming data-driven noise information. Our approach is closely related to \cite{LorenzenTAC2017,HewingLCSS2020}, which employ offline sampling for constraint tightening and probabilistic reachable set computation. Both methods assume a fixed feedback control policy, which preserves convexity in the underlying scenario-based relaxed programs, as required by SO in \cite{Calafiore2005, Campi2008} to enable solutions with a priori guarantees. Disturbance feedback parameterization \cite{Goulart2006} is employed in \cite{Prandini2012} for online SO-based stochastic optimal control, which maintains convexity and is well-suited for short control horizons. In our approach, we design a feedback controller using disturbance samples to ensure the error system remains within a prediction region (PR). Although state-feedback control introduces a nonlinear dependence of dynamics on the feedback gain, we can derive formal guarantees through a two-step training-calibration procedure provided by CP. We define PRs as sublevel sets of nonconformity scores, with thresholds based on empirical quantiles computed from data. To handle joint chance constraints, we introduce nonconformity scores capturing the maximum error across a trajectory. We propose two methods. In the first, we use a training dataset to optimize nonconformity score quantiles for controller design, followed by PR construction via CP on a calibration dataset. In the second method, we first obtain a disturbance PR via CP, and subsequently use the $S$-procedure to design a PR and a feedback gain for the error system. The PRs are used to tighten the constraints, enabling us to solve a deterministic relaxation problem that yields a feedback control policy with probabilistic guarantees. We demonstrate the performance of both methods through a numerical example and compare our approach with the SO-based randomized method in \cite{Prandini2012}.

\section{Preliminaries and problem setup}\label{sec:Prob_setup}

\noindent\textbf{Notation:} The set $\N_{[N_1,N_2]}$ collects all integers in $[N_1,N_2]$. $\bm{x}(a:b)=(x(a),\ldots,x(b))$ is an aggregate vector collecting $x(t)$, $t\in \N_{[a,b]}$, representing a trajectory or a random process. The ceiling operator is $\lceil{}\cdot{}\rceil$. The probability of $Y$ is $\Pr\{Y\}$. $\operatorname{Quantile}(\mathscr D,\delta)$ is the $\delta$-th quantile of a distribution $\mathscr D$, i.e., for $Z \sim \mathscr D$, $\operatorname{Quantile}(\mathscr D,\delta)=\inf\{z:\prb{Z\leq z}\geq \delta\}$. $A$ being a (proper) subset of $B$ is denoted as ($A\subsetneq B$) $A \subseteq B$. The Pontryagin set difference of $S_1, S_2\subseteq \R^n$ is $S_1\ominus S_2 = \{s_1\in \R^n\;|\;s_1  +s_2 \in S_1, \forall s_2 \in S_2 \}$. $\mathrm{int}(X)=\{x\in X\mid \exists e>0,\; \mathbb{B}(e)+x\in X\}$, where $\mathbb{B}(e)=\{x\mid \sqrt{x^\top x}\leq e\}$. The \(i\)th, the smallest, and the largest eigenvalues of a symmetric matrix \(M\) are \(\lambda_i(M)\), \(\lambda_{\min}(M)\), and \(\lambda_{\max}(M)\), respectively. The determinant and trace of $M$ is $\det M$ and $\operatorname{trace} M$.

\noindent\textbf{Conformal Prediction:} Let $\cR^{(0)},\ldots,\cR^{(k)}$ be independent and identically distributed (i.i.d.) random variables. We will refer to $\cR^{(j)}$ as a nonconformity score. Given a failure probability $\theta\in(0,1)$, one wishes to obtain a bound $C\in\R$ for $\cR^{(0)}$ so that 
\begin{equation}\label{eq:prob_for_R0}
    \prb{\cR^{(0)}\leq C} \geq 1-\theta,
\end{equation}
where $C$ is based on the samples $\cR^{(1)},\ldots,\cR^{(k)}$, which we call calibration dataset. Specifically, $C$ may be attained as $C=\quan{\cR^{(1)},\ldots,\cR^{(k)},\infty}{1-\theta}$, which is the $(1-\theta)$th quantile of the empirical distribution $\{\cR^{(1)},\ldots,\cR^{(k)},\infty\}$. Assuming $\cR^{(1)}\leq \cdots \leq \cR^{(k)}$, one can pick $C=\cR^{(p)}$, where $p=\lceil(k+1)(1-\theta)\rceil$, which indicates the $p$th smallest nonconformity score. Note that $C$ is finite with $p\in\N_{[1,k]}$ if $k\geq \lceil(k+1)(1-\theta)\rceil$. This choice of \( C \) ensures that \eqref{eq:prob_for_R0} holds since test point $\cR^{(0)}$ and calibration data $\cR^{(1)}, \ldots, \cR^{(k)}$ are i.i.d. \cite{TibshiraniNeurIPS2019}. This is summarized below. 
\begin{lemma}{\cite[Lemma 1]{TibshiraniNeurIPS2019}}\label{lemma:quantile_lemma}
    If \(\cR^{(0)},\ldots,\cR^{(k)}\) are i.i.d. random variables, then for any $\theta\in(0,1)$, we have 
    \begin{equation}\label{eq:quantile_lemma}
        \Pr\{{\cR^{(0)}}{\leq} {\quan{{\cR^{(1)}{,\ldots,}\cR^{(k)},\infty}}{{1}{-}\theta}}\}{\geq} {1}{-}{\theta}.
    \end{equation}
\end{lemma}
\begin{remark}
    The coverage guarantees in \eqref{eq:quantile_lemma} are marginal as the probability is defined over the randomness in the draw of test and calibration points $\cR^{(0)}$, $\cR^{(1)}, \ldots, \cR^{(k)}$. Conditional coverage guarantees of the form $\Pr\{\cR^{(0)} \leq C \mid \cR^{(1)}, \ldots, \cR^{(k)}\}$ are unfortunately not possible to obtain. However, one can show that the conditional probability is a random variable following a beta distribution centered at $1-\theta$ regardless of $k$ \cite{angelopoulos2022gentle,lindemann2024CPsurvey}. Notably, probably approximately correct coverage guarantees
        $\mathrm{Pr}\{\mathrm{Pr}\{\cR^{(0)}\leq \quan{\cR^{(1)},\ldots,\cR^{(k)},\infty}{1-\hat{\theta}}\}\geq 1-\theta\}\geq 1-\beta$ can be obtained by setting $\hat{\theta}=\theta-\sqrt{\frac{\ln{(\sfrac{1}{\beta})}}{2k}}$, with the ``outer" probability taken with respect to the randomness in the draw of the calibration data $\cR^{(1)},\ldots,\cR^{(k)}$, and $\beta\in(0,1)$ \cite{Vovk2012}.
\end{remark}

\noindent\textbf{Problem Statement:} We wish to solve the chance-constrained optimal control problem 
\begin{subequations}\label{eq:motivation_problem}%
\begin{align}%
   &\operatorname*{Minimize}_{\substack{\bm{u}(0:N-1), \; \bm{x}(1:N)}}
        \mathbb{E}
        \left[
            \sum_{t=0}^{N-1} \ell(x(t), u(t)) + V_f(x(N))
        \right], \label{eq:cost_function_prob} \\
     &\mathrm{s.t.~}  x(t+1) = Ax(t)+Bu(t)+w(t),\; t\in\N_{[0,N)}, \label{eq:dynamics_prob} \\
        &\;\;\;\;\;\; \mathrm{Pr}\{x(t)\in\cX_t, \;\forall t\in\N_{[1,N]}\}\geq 1-\theta,  \label{eq:state_constraints}\\
        &\;\;\;\;\;\; \mathrm{Pr}\{u(t)\in\cU, \;\forall t\in\N_{[0,N-1]}\}\geq 1-\theta, \label{eq:input_constraints}\\
        &\;\;\;\;\;\; x(0)=x_0, \label{eq:initial_condition}
\end{align}%
\end{subequations}%
where $\ell:\R^{n}\times \R^{m}\to \R$ and $V_{f}:\R^n\to \R$ are cost functions penalizing the state and input trajectories $\bm{x}(0:N-1)$, $\bm{u}(0:N-1)$, and the end point $x(N)$, respectively, of the discrete-time linear stochastic system with dynamics in \eqref{eq:dynamics_prob}, and $\theta\in(0,1)$ is a failure probability. In \eqref{eq:dynamics_prob}, \(x(t) \in \R^n\) is a state vector, \(u(t) \in \R^m\) is an input vector, and \(w(t)\) is a disturbance vector consisting of random variables drawn from arbitrary distributions. We assume that $w(t)$, $t\in\N_{[0,N-1]}$, are i.i.d., that is, $w(t)\sim \mathscr{D}$ for all $t\in\N_{[0,N-1]}$, with the underlying multivariate distribution $\mathscr{D}$ being unknown. We assume that $(A,B)$ is stabilizable with $A\in\R^{n\times n}$, $B\in\R^{n\times m}$, and the state and input constraint sets are ellipsoidal regions of $\R^n$ and $\R^m$, respectively, defined as
\begin{subequations}\label{eq:cXtcU}%
    \begin{align}%
        \cX_t&=\{x\in\R^n\mid (x-p_t)^\top P_t (x-p_t) \leq 1\}, \label{eq:cXt}\\
        \cU&=\{u\in\R^m\mid u^\top Q u \leq 1\}, \label{eq:cU}
    \end{align}
\end{subequations}
where $p_t\in\R^n$, $t\in\N_{[1,N]}$, and $P_t\in \R^{n\times n}$, $t\in\N_{[1,N]}$, and $Q\in \R^{m\times m}$ are symmetric positive definite matrices. 

Solving \eqref{eq:motivation_problem} is challenging primarily due to the unknown distribution of \( w(t) \). In the following, we assume that a disturbance dataset, \( \cD^w = \{ \bm{w}^{(0)}, \ldots, \bm{w}^{(k)} \} \), of \( k+1 \) samples of disturbance sequences is available, with $\bm{w}^{(j)}=(w^{(j)}(0),\ldots,w^{(j)}(N-1))$, $j\in\N_{[0,k]}$, and $w^{(j)}(t) \sim \mathscr{D}$ for all $t\in\N_{[0,N-1]}$.

\section{Main results}

Due to linearity in \eqref{eq:dynamics_prob}, the state $x(t)$ can be decomposed into a deterministic part, $z(t)$, and an error, $e(t)$, i.e., $x(t) = z(t) + e(t)$. We will tackle the stochastic control problem in \eqref{eq:motivation_problem}-\eqref{eq:cXtcU} by the control law $u(t)=Ke(t)+v(t)$, where $K\in \R^{m\times n}$ is a state-feedback gain for the pair $(A,B)$, and $v(t)$ is a feedforward control action.
Then, we may write%
\begin{subequations}\label{eq:decomposed_dynamics}%
    \begin{align}%
    z(t+1)&=Az(t)+Bv(t), \label{eq:determ_dyn}\\
    e(t+1)&=\bar{A} e(t)+w(t),\label{eq:error_dyn}
\end{align}    
\end{subequations}
where $z(0)=x(0)$, $e(0)=0$, and $\bar{A}=A+BK$. This standard decomposition technique allows the two systems in \eqref{eq:decomposed_dynamics} to be analyzed independently \cite{CannonTAC2011}. Unlike existing approaches that assume a fixed feedback controller \cite{HewingLCSS2020, HewingCDC2018}, we aim to design both feedback and feedforward control terms using the available disturbance dataset $\cD^w$. We partition $\cD^w$ into training and calibration datasets, $\cD^w_\mathrm{train} = \{\bm{w}^{(k_1+1)}, \ldots, \bm{w}^{(k)}\}$ and $\cD^w_\mathrm{cal} = \{\bm{w}^{(0)}, \ldots, \bm{w}^{(k_1)}\}$, where $k_1 + 1 < k$. The training dataset is used to compute design parameters, such as \(K\), while the calibration dataset is used to generate empirical distributions of nonconformity scores defined over error and input trajectories. Using Lemma \ref{lemma:quantile_lemma}, we then derive prediction regions for the error state \(e(t)\) and the feedback term \(Ke(t)\), with formal guarantees. Next, we define prediction regions for random vectors, and subsequently, show how to solve a relaxation of the problem in \eqref{eq:motivation_problem} optimizing over control actions, given a fixed feedback gain $K$.

\subsection{Open-loop control}

\begin{definition}
     Let $\xi\in\R^n$ be a random vector. We call $\mathscr{E}_{1-\theta}(\xi)\subseteq \R^n$ a prediction region (PR) for $\xi$ at a confidence level $1-\theta$, if $\mathrm{Pr}\{\xi\in \mathscr{E}_{1-\theta}(\xi)\}\geq 1-\theta$. We also denote $\mathscr{E}_{1-\theta}^{a:b}(\xi(t))$ a prediction region for the random process $(\xi(a),\ldots,\xi(b))$, if $\mathrm{Pr}\{\xi(t)\in \mathscr{E}_{1-\theta}^{a:b}(\xi(t))\; \forall t\in \N_{[a,b]}\}\geq 1-\theta$. 
\end{definition}

Given PRs $\mathscr{E}_{1-\theta}^{1:N}(e(t))$, $\mathscr{E}_{1-\theta}^{0:N-1}(Ke(t))$, we can tighten the constraints in \eqref{eq:cXtcU} to produce constraints for the deterministic variables $z(t)$, $v(t)$, in \eqref{eq:determ_dyn}. Conditions for tightening of the constraints in \eqref{eq:cXtcU} are given next.

\begin{lemma}\label{lemma:conditions}
    Consider the ellipsoidal constraints in \eqref{eq:cXtcU}, and the error system in \eqref{eq:error_dyn}. Let $\mathscr{E}_{1-\theta}^{1:N}(e(t))$ and $\mathscr{E}_{1-\theta}^{0:N-1}(Ke(t))$ be prediction regions for $(e(1),\ldots,e(N))$ and $(Ke(0),\ldots,Ke(N-1))$, respectively. If $\mathscr{E}_{1-\theta}^{1:N}(e(t)) \subseteq \mathbb{B}(C_e)$ and $\mathscr{E}_{1-\theta}^{0:N-1}(Ke(t)) \subseteq \mathbb{B}(C_u)$, with $C_e\in(0, \sfrac{1}{\sqrt{\lambda_\mathrm{max}(P_t)}})$, $\forall t\in\N_{[1,N]}$, and $C_u\in(0,\sfrac{1}{\sqrt{\lambda_\mathrm{max}(Q)}})$, respectively, then
\begin{subequations}\label{eq:tightening_conditions}
    \begin{align}
    &\mathrm{int}\left(\cX_t \ominus \mathscr{E}_{1-\theta}^{1:N}(e(t)) \right) \neq \emptyset \; \forall t\in\N_{[1,N]},\label{eq:constraint_BCe} \\
    &\mathrm{int}\left(\cU \ominus \mathscr{E}_{1-\theta}^{0:N-1}(Ke(t))\right) \neq \emptyset. \label{eq:constraint_BCu}
\end{align}%
\end{subequations}%
\end{lemma}%
\begin{proof}%
Define $c_{\min}=\min_t \sfrac{1}{\sqrt{\lambda_{\max}(P_t) }}$. For the condition in \eqref{eq:constraint_BCe}, it suffices to show that \(\mathrm{int}(\hat{\cX}_t \ominus \mathbb{B}(C_e)) \neq \emptyset\), $\forall t\in\N_{[1,N]}$, where $\hat{\cX}_t=\cX_t-p_t$. Note that $\hat{\cX}_t\ni 0$ is an ellipsoid, with the length of its smallest semi-axis equal to \(\sfrac{1}{\sqrt{\lambda_{\max}(P_t)}}\). Thus, $\mathbb{B}(c_{\min})\subseteq \hat{\cX}_t$, $\forall t\in\N_{[1,N]}$. Moreover, since $C_e\in(0,c_{\min})$, we have $\mathbb{B}(c_{\min})\ominus \mathbb{B}(C_e)=\mathbb{B}(c_{\min}-C_e)\subseteq \hat{\cX}_t \ominus \mathbb{B}(C_e)$, $\forall t\in\N_{[1,N]}$. Taking the interior on both sides of this inclusion, the result follows, i.e., $\mathrm{int}(\mathbb{B}(c_{\min}-C_e)) \subseteq \mathrm{int}(\hat{\cX}_t\ominus \mathbb{B}(C_e))\neq \emptyset$, since $\mathrm{int}(\mathbb{B}(c_{\min}-C_e))\neq \emptyset$. The condition in \eqref{eq:constraint_BCu} is proved similarly.
\end{proof}

We now present a relaxation of the problem in \eqref{eq:motivation_problem} with tighter state and input constraints based on PRs for a fixed $K$ similar to the approaches in \cite{CannonTAC2011, LorenzenTAC2017, HewingCDC2018}.

\begin{theorem}\label{thm:determ_prob}
Consider the optimization problem in \eqref{eq:motivation_problem}-\eqref{eq:cXtcU}, the system in \eqref{eq:decomposed_dynamics} with PRs, $\mathscr{E}_{1-\theta}^{1:N}(e(t))\subsetneq \R^n$, $\mathscr{E}_{1-\theta}^{0:N-1}(Ke(t))\subsetneq \R^m$, for a fixed gain $K$, and 
\begin{subequations}\label{eq:determ_problem}
\begin{align}
   &\operatorname*{Minimize}_{\substack{\bm{v}(0:N-1), \; \bm{z}(1:N)}}
            \sum_{t=0}^{N-1}\ell(z(t), v(t)) + V_f(z(N)) \label{eq:cost_function_determ} \\
     &\mathrm{s.t.~}  z(t+1) = Az(t)+Bv(t),\; t\in\N_{[0,N)}, \label{eq:determ_dynamics} \\
        &\;\;\;\;\;\; z(t)\in\cZ_t, \;\forall t\in\N_{[1,N]},  \label{eq:determ_state_constraints}\\
        &\;\;\;\;\;\; v(t)\in\cV, \;\forall t\in\N_{[0,N-1]}, \label{eq:determ_input_constraints}\\
        &\;\;\;\;\;\; z(0)=x_0, \label{eq:determ_initial_condition}
\end{align}    
\end{subequations}
where $\cZ_t=\cX_t\ominus \mathscr{E}_{1-\theta}^{1:N}(e(t))$, $t\in\N_{[1,N]}$, and $\cV=\cU\ominus \mathscr{E}_{1-\theta}^{0:N-1}(Ke(t))$. Assume that the conditions in \eqref{eq:tightening_conditions} are true, and that \eqref{eq:determ_problem} is feasible for $z(0)=x_0$, with an optimal solution $(\bm{v}^\ast(0:N-1), \bm{z}^\ast(1:N))$. Let $\bm{u}(0:N-1)=(u(0),\ldots,u(N-1))$, with $u(t)=Ke^\ast(t)+v^\ast(t)$, where $e^\ast(t)=x(t)-z^\ast(t)$, $t\in\N_{[0,N-1]}$, and $x(t)$ is the state of the system \eqref{eq:dynamics_prob}. Let the trajectory of \eqref{eq:dynamics_prob}, $\bm{x}(1:N)$, originated at $x_0$ be driven by input and disturbance sequences, $\bm{u}(0:N-1)$, $\bm{w}(0:N-1)$, where individual disturbances follow the same distribution as the elements of $\cD^w_\mathrm{cal}$. Then, $(\bm{u}(0:N-1),\bm{x}(1:N))$ is a feasible solution to \eqref{eq:motivation_problem}.    
\end{theorem}
\begin{proof}
Since the conditions in \eqref{eq:tightening_conditions} are true, it suffices to show that the probabilistic constraints in \eqref{eq:state_constraints}-\eqref{eq:input_constraints} are feasible. Define events $X:= x(t)\in\cX_t \; \forall t\in\N_{[1,N]}$, $E:= e(t)\in \mathscr{E}_{1-\theta}^{1:N}(e(t))\; \forall t\in\N_{[1,N]}$, and $Z:= z(t)\in \cZ_t\; \forall t\in\N_{[1,N]}$, and let $\hat{E}$ be the complement of $E$. Since \eqref{eq:determ_problem} is feasible by assumption, $\prb{Z}=1$, implying that  $\prb{X \mid E}=1$, by definition of $\cZ_t=\cX_t\ominus\mathscr{E}_{1-\theta}^{1:N}(e(t))$. From the law of total probability, we have $\prb{X}=\prb{X\mid E}\prb{E}+\prb{X\mid \hat{E}}\prb{\hat{E}}\geq 1-\theta$, which follows from the fact that $\prb{X\mid E}=1$, $\prb{E}\geq 1-\theta$, and $\prb{X\mid \hat{E}}\prb{\hat{E}}\geq 0$. The constraint in \eqref{eq:input_constraints} can be shown by similar reasoning. 
\end{proof}

We note that the conditions in \eqref{eq:tightening_conditions} of Lemma \ref{lemma:conditions} ensure that \eqref{eq:determ_problem} is well defined. Theorem \ref{thm:determ_prob} can be stated independently. We also remark that the input $u(t) = Ke(t) + v(t)$ may be unbounded as $e(t)$ is influenced by a stochastic, potentially unbounded, disturbance. The chance constraint in \eqref{eq:input_constraints} allows the input constraint to be violated for some disturbance realizations, with the parameter $\theta$ specifying the extent of such violations. Next, we design PRs that bound the feedback term $Ke(t)$ with a probability of at least $1-\theta$. 

\subsection{Prediction regions and feedback control}\label{sec:fb_design}

Assuming that PRs for the error system in \eqref{eq:error_dyn} with a fixed state-feedback gain \( K \) are available, Theorem \ref{thm:determ_prob} provides a method to construct a feasible solution for the problem in \eqref{eq:motivation_problem}–\eqref{eq:cXtcU} by solving a tractable optimization problem. Here, we outline two methods for designing \( K \) and approximate PRs for the error system. The first method formulates an optimization problem over a training dataset of error trajectories, where each trajectory sample depends nonlinearly on \( K \). After solving this, conformal prediction is applied on a calibration dataset to compute PRs, referred to as the \textit{direct method}, as CP is applied directly to error trajectories. The second method uses conformal prediction on disturbance samples to identify an ellipsoidal set that bounds disturbance sequences with the specified probability. The \( S \)-procedure is then used to compute an admissible state-feedback gain and an ellipsoidal PR. This is called the \textit{indirect method}, as CP is applied to disturbance samples to indirectly infer coverage for the error system's PRs.

\subsubsection{Direct method}

From the available disturbance dataset $\cD^w$, we define a trajectory dataset as
\begin{subequations}\label{eq:cal_traj}
    \begin{align}
        \cD^e&=\{\bm{e}^{(0)},\ldots,\bm{e}^{(k)}\}, \label{eq:cDe_cal} \\
        \bm{e}^{(j)}&=(e^{(j)}(1),\ldots,e^{(j)}(N)), \; j\in\N_{[0,k]} \label{eq:bmej}\\
        e^{(j)}(t)&=\sum_{i=1}^t(A+BK)^{i-1}w^{(j)}(t-i), \; t\in\N_{[1,N]}. \label{eq:ejt}
    \end{align}
\end{subequations}
In the following, we partition $\cD^e$ into $\cD^e_{\mathrm{train}}$ and $\cD^e_{\mathrm{cal}}$, where trajectory samples in $\cD^e_{\mathrm{train}}$ are constructed by disturbance samples in $\cD^w_{\mathrm{train}}$ and are parameterized by the feedback gain $K$, while trajectory samples in $\cD^e_{\mathrm{cal}}$ are constructed by disturbance samples in $\cD^w_{\mathrm{cal}}$ for a fixed $K$. 
We also introduce the nonconformity scores 
\begin{subequations}\label{eq:nonconf_scores}
    \begin{align}
    \cR^{(j)}_e&=\max(\|e^{(j)}(1)\|,\ldots,\|e^{(j)}(N)\|), \label{eq:cRe} \\
    \cR^{(j)}_u&=\max(\|Ke^{(j)}(0)\|,\ldots,\|Ke^{(j)}(N-1)\|), \label{eq:cRu}
    \end{align}
\end{subequations}
which will help us identify PRs for the random processes $(e(1),\ldots,e(N))$ and $(Ke(0),\ldots,Ke(N-1))$, as Euclidean balls bounding $\|e(t)\|$ and $\|Ke(t)\|$ uniformly for all $t \in \mathbb{N}_{[0,N]}$, enabling efficient tightening of the ellipsoids in \eqref{eq:cXtcU}. Consider now the optimization problem
\begin{subequations}\label{eq:Kproblem}
\begin{align}
   &\operatorname*{Minimize}_{K, \; \eta_e,\; \eta_u} \eta_e  + \gamma \eta_u \label{eq:cost_function_Kproblem1} \\
     &\mathrm{s.t.~}
     \prb{\cR^{(0)}_e \leq \eta_e} \geq 1-\theta,  \label{eq:constraintCe_Kproblem} \\
     & \;\;\;\;\;\; \prb{\cR^{(0)}_u \leq \eta_u} \geq 1-\theta , \label{eq:constraintCu_Kproblem}
\end{align}
\end{subequations}
where the objective is to obtain a state-feedback gain \(K\) and minimal PRs $\mathscr{E}_{1-\theta}^{1:N}(e(t))$, $\mathscr{E}_{1-\theta}^{0:N-1}(Ke(t))$ induced by the constraints in \eqref{eq:constraintCe_Kproblem}-\eqref{eq:constraintCu_Kproblem}, and the nonconformity scores \(\cR^{(j)}_e\), \(\cR^{(j)}_u\) in \eqref{eq:nonconf_scores}. The objective in \eqref{eq:cost_function_Kproblem1} balances the trade-off between minimizing the error cost \( \eta_e \) and the feedback control cost \( \eta_u \), as indicated by the nonconformity scores \( \cR^{(0)}_e \) and \( \cR^{(0)}_u \) in \eqref{eq:nonconf_scores}. A specific value for the tuning parameter \( \gamma \), which leads to a cost function that supports the desired performance while adhering to state and input constraints, is recommended later. At first glance, one might consider using Lemma \ref{lemma:quantile_lemma} to replace the probabilistic constraints in \eqref{eq:constraintCe_Kproblem}-\eqref{eq:constraintCu_Kproblem} with constraints as in \eqref{eq:quantile_lemma}, that is, $\quan{\cR_e^{(1)},\ldots,\cR_e^{(k_1)},\infty}{1-\theta}\leq \eta_e$, and $\quan{\cR_u^{(1)},\ldots,\cR_u^{(k_1)},\infty}{1-\theta}\leq \eta_u$, by constructing empirical nonconformity scores $\cR_e^{(j)}$ and $\cR_u^{(j)}$ using the available calibration dataset $\cD^e_\mathrm{cal}$. However, this is not straightforward because the random variables $\cR_e^{(0)},\ldots,\cR_e^{(k_1)}$ and $\cR_u^{(0)},\ldots,\cR_u^{(k_1)}$ do not retain the i.i.d property, as they depend on the decision variable $K$. Instead, we can first obtain a state-feedback gain $K$ using a separate training dataset $\cD^e_\mathrm{train}$ and subsequently apply conformal prediction to attain coverage guarantees for the constraints in \eqref{eq:constraintCe_Kproblem}-\eqref{eq:constraintCu_Kproblem} by computing a calibration dataset $\cD^e_\mathrm{cal}$ for the obtained gain $K$. This approach is shown next.

Let $\cD^e_\mathrm{train}=\{\bm{e}^{(k_1+1)},\ldots, \bm{e}^{(k)}\}$, where $\bm{e}^{(j)}$ is a trajectory sample parameterized over state-feedback gains $K\in\R^{m\times n}$ and driven by a training disturbance sequence $\bm{w}^{(j)}\in\cD^w_\mathrm{train}$, with $j\in\N_{[k_1+1,k]}$. Then, we formulate the problem of finding a state-feedback gain $K$ by solving
\begin{subequations}\label{eq:Ktrain}
\begin{align}
   &\operatorname*{Minimize}_{K, \; \eta_e < \eta_{e}^{\max},\; \eta_u < \eta_{u}^{\max}} \eta_e  + \gamma \eta_u \label{eq:cost_function_Kproblem} \\
     &\mathrm{s.t.~} \quan{\cR_e^{(k_1+1)}(K),\ldots,\cR_e^{(k)}(K)}{\hat{\theta}} \leq  \eta_e, \label{eq:constraintCe_Ktrain} \\
     & \;\;\;\;\;\; \quan{\cR_u^{(k_1+1)}(K),\ldots,\cR_u^{(k)}(K)}{\hat{\theta}} \leq  \eta_u, \label{eq:constraintCu_Ktrain} 
\end{align}
\end{subequations}
where 
$\eta_e^{\max} = \min_{t\in\N_{[1,N]}} \left(\sfrac{1}{\sqrt{\lambda_{\mathrm{max}}(P_t)}}\right)$, 
$\eta_u^{\max} = \sfrac{1}{\sqrt{\lambda_{\mathrm{max}}(Q)}}$,
$\cR_{e}^{(j)}(K)$ ($\cR_{u}^{(j)}(K)$) denotes the parameterization over the feedback gain $K$, 
and we choose $\gamma = \tfrac{\eta_e^{\max}}{\eta_u^{\max}}$, and $\hat{\theta}=(1+\frac{1}{k-k_1-1})(1-\theta)$. Note that $\infty$ is omitted from the empirical distributions in \eqref{eq:constraintCe_Ktrain}-\eqref{eq:constraintCu_Ktrain} and $1-\theta$ is replaced by $(1+\frac{1}{k-k_1-1})(1-\theta)$ (see \cite[Sec. 2.1]{lindemann2024CPsurvey} for details). Solving \eqref{eq:Ktrain} is indeed difficult due to the nonlinear dependence of the constraints in \eqref{eq:constraintCe_Ktrain}-\eqref{eq:constraintCu_Ktrain} on the decision variable $K$. One can employ a nonlinear solver to obtain a feasible solution to \eqref{eq:Ktrain}. See, e.g., Sec. \ref{sec:example}, where we employ a genetic algorithm to solve \eqref{eq:Ktrain} for an academic example.

Next, given a state-feedback gain $K$ obtained from \eqref{eq:Ktrain}, we can provide PRs, $\mathscr{E}_{1-\theta}^{1:N}(e(t))$, $\mathscr{E}_{1-\theta}^{0:N-1}(Ke(t))$, via conformal prediction using empirical distributions of the nonconformity scores in \eqref{eq:nonconf_scores} generated by constructing the calibration dataset $\cD^e_\mathrm{cal}$. This is formally stated next.

\begin{lemma}\label{lemma:prediction_regions}
Construct the calibration trajectory dataset $\cD^e_\mathrm{cal}$ as in \eqref{eq:cal_traj}, using the calibration disturbance set $\cD^w_\mathrm{cal}$ and the solution \( K \) from \eqref{eq:Ktrain}. Let $\{\cR_e^{(0)},\ldots, \cR_e^{(k_1)}\}$, $\{\cR_u^{(0)},\ldots,\cR_u^{(k_1)}\}$ be the empirical distributions in \eqref{eq:cRe}, \eqref{eq:cRu}, respectively, computed by the samples in $\cD^e_\mathrm{cal}$ and the gain \( K \). Compute $C_e=\quan{\cR_e^{(1)},\ldots, \cR_e^{(k_1)},\infty}{1-\theta}$ and $C_{Ke}=\quan{\cR_u^{(1)},\ldots, \cR_u^{(k_1)},\infty}{1-\theta}$. Then, if $\mathscr{E}_{1-\theta}^{1:N}(e(t)) \coloneqq \mathbb{B}(C_e)$, $\mathscr{E}_{1-\theta}^{0:N-1}(Ke(t)) \coloneqq \mathbb{B}(C_{Ke})$, we have
\begin{subequations}\label{eq:objective1}
    \begin{align}
    &\Pr\left\{ e^{(0)}(t) {\in} \mathscr{E}_{1-\theta}^{1:N}(e(t)),  t\in\N_{[1,N]}  \right\}\geq 1{-}\theta, \label{eq:constraint_e} \\
    &\Pr\left\{ Ke^{(0)}(t) {\in} \mathscr{E}_{1-\theta}^{0:N-1}(Ke(t)),  t{\in}\N_{[0,N)} \right\}\geq 1{-}\theta. \label{eq:constraint_Ke}
\end{align}  
\end{subequations}
\end{lemma}%
\begin{proof}
First we prove that $\cD^e_\mathrm{cal}$ consists of $k_1+1$ i.i.d. random trajectories of \eqref{eq:error_dyn}. For each fixed \( t\in\N_{[1,N]}\), the random vector \( e^{(j)}(t) \) in \eqref{eq:ejt} is identically distributed across all \( j \in\N_{[0,k_1]} \). This means that for any fixed time index \( t \), the distribution of \( e^{(j)}(t) \) does not depend on the index \( j \) for a given $K$. Therefore, since the distribution of each individual \( e^{(j)}(t) \) is the same for all \( j \), it follows that the entire random process \( \bm{e}^{(j)}(1:N) \) in \eqref{eq:bmej} is identically distributed across \( j \in\N_{[0,k_1]} \). By the independence of $\bm{w}^{(j)}$ across all $j\in\N_{[0,k_1]}$, \( \bm{e}^{(j)}(1:N) \), $j\in\N_{[0,k_1]}$, are i.i.d. random processes, implying that $\cR_e^{(0)},\ldots,\cR_e^{(k_1)}$ are $k_1+1$ i.i.d random variables. Thus, by Lemma \ref{lemma:quantile_lemma}, we obtain the coverage $\prb{\cR_e^{(0)} \leq C_e} \geq 1-\theta$, which can be written as $\prb{\max(\|e^{(0)}(1)\|, \ldots, \|e^{(0)}(N)\|) \leq C_e} \geq 1-\theta$ by \eqref{eq:cRe}, or as $\prb{e^{(0)}(1) \in \mathbb{B}(C_e) \land \cdots \land e^{(0)}(N) \in \mathbb{B}(C_e)} \geq 1-\theta$, leading to \eqref{eq:constraint_e}. The proof of \eqref{eq:constraint_Ke} follows identical lines. 
\end{proof}

We summarize important remarks for selecting the nonconformity scores in \eqref{eq:nonconf_scores} as follows: 1) The PRs obtained in Lemma \ref{lemma:prediction_regions} give marginal guarantees, that is, the probabilities in \eqref{eq:objective1} are averaged over the randomness of both the training and calibration data. 2) The ball $\mathbb{B}(\|K\|C_e )$ is a conservative PR, i.e., $\mathscr{E}_{1-\theta}^{0:N-1}(Ke(t))$, motivating the use of two nonconformity scores. 3) We can select \(\infty\)-norm-based nonconformity scores to infer box-shaped PRs when more appropriate, e.g., for polyhedral constraints. 

\subsubsection{Indirect method}

First, we compute an ellipsoid containing all the training disturbance samples in $\cD^w_\mathrm{train}$ by solving the optimization
\begin{subequations}\label{eq:distur_ellipsoid}
\begin{align}
   &\operatorname*{Minimize}_{\hat{Y}\succ 0} \log \det \hat{Y}^{-1} \label{eq:cost_function_min_vol_ellipsoid} \\
     &\mathrm{s.t.~} \|\hat{Y} w^{(j)}(t)\|\leq 1, \; t\in\N_{[0,N-1]}, \; j\in\N_{[k_1+1,k]}.\label{eq:constraints_min_vol_ellipsoid}
\end{align}    
\end{subequations}
Note that \eqref{eq:distur_ellipsoid} is a convex problem \cite[Sec. 8.4]{boyd2004convex}, with $\hat{Y}$ being a symmetric positive definite matrix. Also, the volume of $\hat{\cW}=\{w\mid \|\hat{Y}w\|\leq 1\}$ is minimum since it is proportional to $\det \hat{Y}$ by $\hat{Y}\succ 0$. Next, we obtain an ellipsoidal prediction region $\mathscr{E}_{1-\theta}^{0:N-1}(w(t))$. Consider the nonconformity score
\begin{equation}\label{eq:cRw}
    \cR^{(j)}_w=\max(\|\hat{Y}w^{(j)}(0)\|,\ldots,\|\hat{Y}w^{(j)}(N-1)\|),
\end{equation}
and compute 
\begin{equation}\label{eq:Cw}
    C_w = \quan{\cR_w^{(1)},\ldots,\cR_w^{(k_1)},\infty}{1-\theta},
\end{equation}
over the calibration dataset $\cD^w_{\mathrm{cal}}$. Then, by Lemma \ref{lemma:quantile_lemma}, we obtain $\prb{\cR^{(0)}_w \leq C_w}\geq 1-\theta$, which implies that $\prb{\|\hat{Y}w^{(0)}(t)\|\leq C_w \; \forall t\in\N_{[0,N-1]}}\geq 1-\theta$, i.e., the ellipsoid 
\begin{equation}\label{eq:cW}
    \cW = \{w\mid w^{\top} Y w \leq 1\},
\end{equation}
is a PR, $\mathscr{E}_{1-\theta}^{0:N-1}(w(t))$, where $Y = \hat{Y}^{\top} \hat{Y}/C_w^2$. Thus, the feedback design problem can be reduced to the synthesis of an ellipsoidal region for the error system \eqref{eq:error_dyn} that is robustly controlled invariant \cite[Def. 2.3]{Blanchini1999invariance} by a feedback controller $K$ for all disturbances lying in $\cW$ in \eqref{eq:cW}. In other words, we seek 
\begin{equation}\label{eq:cE}
    \cE=\{e\mid e^\top \Phi e \leq 1\},
\end{equation}
where $\Phi\succ 0$ is a symmetric positive definite matrix, and a state-feedback gain $K$, such that 
\begin{equation}\label{eq:invarianceK} 
    [(A+BK)e+w]^\top \Phi [(A+BK)e+w] \leq 1, \; \forall w\in\cW,
\end{equation}
which ensures that $\cE$ in \eqref{eq:cE} is invariant for all disturbance sequences in $\cW$, thereby serving as a PR. By $S$-procedure \cite[Appendix B.2]{boyd2004convex}, and the fact that $\Phi\succ 0$ and $Y\succ 0$ \cite[Theorem 4.2]{Polik2007}, the invariance condition in \eqref{eq:invarianceK} can be translated into finding a feasible solution to the BMI:
\begin{subequations}\label{eq:LMIs}
    \begin{align}
        \begin{bmatrix} \lambda_0\Phi -\bar{A}^{\top} \Phi \bar{A} & -\bar{A}^{\top} \Phi \\ -\Phi \bar{A} & \lambda_1 Y -\Phi  \end{bmatrix} \succeq & 0,  \label{eq:LMI1_Phi}\\
       \lambda_0, \lambda_1 \geq 0,\; 1-\lambda_0 -\lambda_1 \geq& 0, \label{eq:lambda_1}
    \end{align}
\end{subequations}
where $\bar{A}=A+BK$. By applying Schur complement twice and noting that $\lambda_1Y - \Phi\succ 0$ is equivalent to $\Phi^{-1}-\sfrac{1}{\lambda_1}Y^{-1}\succ 0$, \eqref{eq:LMI1_Phi} can be enforced by requiring
\begin{equation}\label{eq:LMI_phi_hat_psi}
    \begin{bmatrix} \hat{\Phi}-\sfrac{1}{\lambda_1}Y^{-1} & A\hat{\Phi}+B\Psi \\ (A\hat{\Phi}+B\Psi)^\top & \lambda_0 \hat{\Phi}\end{bmatrix}\succeq 0,
\end{equation}
which is linear in $\hat{\Phi}$ and $\Psi$, where $\hat{\Phi} = \Phi^{-1}$ and $\Psi = K\hat{\Phi}$. To ensure that 
$K=\Psi\hat{\Phi}^{-1}$ is an admissible linear controller, we also require that $\max_{e^\top \Phi e\leq 1} \|Q^{\sfrac{1}{2}}Ke\|< 1$, which, by using the state-space transformation $\hat{e}=\Phi^{\sfrac{1}{2}}e$, is written as 
\begin{equation}\label{eq:input_LMI}
    \begin{bmatrix}
        \hat{\Phi} & \Psi^\top Q^{\sfrac{1}{2}}\\ Q^{\sfrac{1}{2}}\Psi & I
    \end{bmatrix}\succ 0.
\end{equation}
We are now ready to state the following result.
\begin{theorem}\label{thm:Spro}
    Consider the ellipsoid $\cW$ as defined in \eqref{eq:cW} such that $\prb{w^{(0)}(t)\in\cW \; \forall t\in\N_{[0,N-1]}}\geq 1-\theta$. Let $\hat{\Phi}^\ast$, $\Psi^\ast$, $\lambda_0^\ast$, $\lambda_1^\ast$ be a feasible solution to the following program:
    \begin{subequations}\label{eq:optimizaton_thm2}
    \begin{align}
    &\operatorname*{Minimize}_{\hat{\Phi}\succ 0,\;\Psi, \; \lambda_0,\; \lambda_1 } \operatorname{trace} \hat{\Phi} \label{eq:opt_thm2_cost_function} \\
     &\mathrm{s.t.~} \eqref{eq:lambda_1}, \; \eqref{eq:LMI_phi_hat_psi},\; \eqref{eq:input_LMI}, \; \hat{\Phi} \prec (P_t)^{-1},\; \forall t\in\N_{[1,N]} \label{eq:constraints_opt_thm2}
\end{align}    
\end{subequations}
    and define $\Phi=(\hat{\Phi}^\ast)^{-1}$. Consider the error system in \eqref{eq:error_dyn}, with $K=\Psi^\ast \Phi$, and the ellipsoidal regions in \eqref{eq:cXtcU}, and define $\cE=\{e\mid e^\top \Phi e \leq 1\}$. Then,
    i) $\prb{e^{(0)}(t)\in \cE \; \forall t\in\N_{[1,N]}} \geq 1-\theta$,
    ii) $\mathrm{int}(\cX_t \ominus \cE) \neq \emptyset \; \forall t\in\N_{[1,N]}$,
    iii) $\mathrm{int}(\cU \ominus \cE_u) \neq \emptyset$, where $\cE_u=\{u=Ke\mid e\in \cE\}$.   
\end{theorem}
\begin{proof}
    i) By the feasibility of \eqref{eq:optimizaton_thm2}, the ellipsoid $\cE$ is robustly invariant for \eqref{eq:error_dyn} for all $w(t)\in\cW$ when $K=\Psi^\ast \Phi$, that is $e(t)\in \cE$ for all $w(t)\in\cW$. Since $\prb{w^{(0)}(t)\in\cW\; \forall t\in\N_{[0,N-1]}}\geq 1-\theta$, one can deduce that $\prb{e^{(0)}(t)\in\cE\; \forall t\in\N_{[1,N]}}\geq 1-\theta$. ii) Without loss of generality, assume that $\cX_t$ is centered at the origin. By the feasibility of the constraints in \eqref{eq:constraints_opt_thm2}, we have that $\Phi\succ P_t$, for $t\in\N_{[1,N]}$, which implies that $\cE\subsetneq \mathrm{int}(\cX_t)$, $\forall t\in\N_{[1,N]}$. Then, it follows that $\mathrm{int}(\cX_t\ominus \cE)\neq \emptyset$. iii) Similarly, by the feasibility of the LMI in \eqref{eq:input_LMI}, it follows that $\cE_u\subsetneq \mathrm{int}(\cU)$, which implies that $\mathrm{int}(\cU\ominus \cE_u)\neq \emptyset$.  
\end{proof}

Note that the optimization in \eqref{eq:optimizaton_thm2} is convex for any fixed pair \((\lambda_0, \lambda_1)\) and thus can be solved using a grid search over \((\lambda_0, \lambda_1)\). It also minimizes an upper bound on \(\det \hat{\Phi}\), as shown by \(\det \hat{\Phi} \leq \left(\frac{1}{n} \operatorname{trace} \hat{\Phi}\right)^n\).

\section{Example}\label{sec:example}

We consider the problem in \eqref{eq:motivation_problem}-\eqref{eq:cXtcU} for a discrete-time double integrator system, with a horizon of $N=100$ time steps, cost functions $\ell(x(t),u(t))=(u(t))^2$, $V_f(x(100))=100x(100)^\top x(100)$, failure probability $\theta=0.05$, $A={\tiny\begin{bmatrix}
    1& 0.5 \\0 & 1
\end{bmatrix}}$, $B={\tiny\begin{bmatrix}
    0\\ 0.5
\end{bmatrix}}$, and initial condition $x(0)=(2,-1)$. For the constraints in \eqref{eq:state_constraints}-\eqref{eq:input_constraints}, we select $P_t=I_2\sfrac{1}{10}$, $p_t=0$, for  $t\in\N_{[1,N]}$, and $Q=1$. For the unknown vector \( w(t) = (w_1(t), w_2(t)) \), we generate calibration and training disturbance datasets, \(\cD^w_{\mathrm{train}}\) and \(\cD^w_{\mathrm{cal}}\), each with 100 sequence samples. We sample \( w_1(t) \) from \(\mathcal{N}(-0.01, 0.005)\), and \( w_2(t) \) from a gamma distribution with shape 5.5 and scale 0.005, each multiplied equally likely by \(1\) or \(-1\). 

\par
\begin{figure}[]
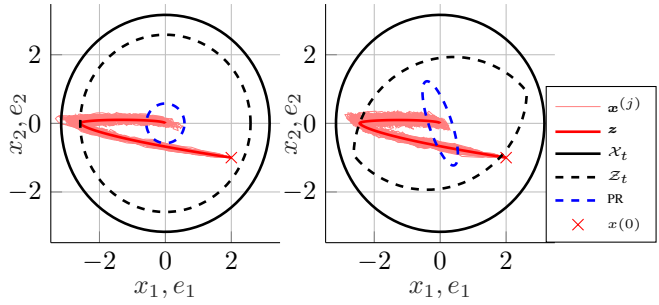
  
  \centering
  \begin{subfigure}[]{0.35\columnwidth} 
    \centering
    \input{figures/fig1}
  \end{subfigure}
  \hfill
  \begin{subfigure}[]{0.6\columnwidth} 
    \centering
    \input{figures/fig2}
  \end{subfigure}
  \caption{\textbf{Direct method} (left), \textbf{Indirect method} (right): Representation of the original state constraint $\cX_t$ (solid black), the tighter state constraint $\cZ_t=\cX_t\ominus \mathscr{E}^{1:N}_{0.95}(e(t))$ (dashed black), the PR $\mathscr{E}^{1:N}_{0.95}(e(t))$ (dashed blue), 100 sample trajectories $\bm{x}^{(j)}(0:N)$ (light red), and the deterministic trajectory $\bm{z}(0:N)$ (red), with initial condition $x(0)$ (red cross).}
  \label{fig:PRs}
\end{figure}

\textbf{Direct method:} We formulate the problem in \eqref{eq:Ktrain}, where $\gamma=\sfrac{\eta_e^{\max}}{\eta_u^{\max}}$, with $\eta_e^{\max}=\sqrt{10}$, $\eta_u^{\max}=1$, and the constraints in \eqref{eq:constraintCe_Ktrain}-\eqref{eq:constraintCu_Ktrain} are constructed using the training dataset $\cD^w_{\mathrm{train}}$. To solve this problem we employ a genetic algorithm (GA), with a population size of 150 candidates for 50 generations. The GA returns $K_{\mathrm{dir}}={\tiny \begin{bmatrix}
    -0.241 & -0.787
\end{bmatrix}}$ within 5 minutes. Next, we obtain PRs, $\mathscr{E}_{0.95}^{1:N}(e(t))$, $\mathscr{E}_{0.95}^{0:N-1}(Ke(t))$, by constructing a calibration trajectory dataset $\cD^e_{\mathrm{cal}}$ by the disturbance dataset $\cD^w_{\mathrm{cal}}$ as in \eqref{eq:cal_traj} for the obtained $K_{\mathrm{dir}}$. Following Lemma \ref{lemma:prediction_regions}, we obtain $\mathscr{E}_{1-\theta}^{1:N}(e(t))=\mathbb{B}(0.5785)$ (see blue dashed circle in Fig. \ref{fig:PRs} (left)) and $\mathscr{E}_{1-\theta}^{0:N-1}(Ke(t))=\mathbb{B}(0.1271)$, which verify the conditions in \eqref{eq:tightening_conditions} of Lemma \ref{lemma:conditions}. 

\par \textbf{Indirect method:} We construct an ellipsoid \( \hat{\mathcal{W}} \) that confines all training disturbance samples in \( \mathcal{D}^w_{\mathrm{train}} \) by solving the problem in \eqref{eq:distur_ellipsoid}. Then, using the nonconformity score from \eqref{eq:cRw} and computing the empirical quantile from \eqref{eq:Cw} over the calibration dataset \( \mathcal{D}^w_{\mathrm{cal}} \) we obtain a tighter ellipsoid \( \mathcal{W} \subseteq \hat{\mathcal{W}}, \) defined in \eqref{eq:cW} with $Y={\tiny{\begin{bmatrix}
    12.6733 &  -1.0720\\
   -1.0720 & 114.7949
\end{bmatrix}}}$, which is a PR, \( \mathscr{E}_{1-\theta}^{0:N-1}(w(t)) \), by Lemma \ref{lemma:prediction_regions}. For the obtained PR, $\cW$, we construct a feedback gain $K_{\mathrm{ind}}={\tiny \begin{bmatrix}
  -1.4140  & -2.3412
\end{bmatrix}}$, and the ellipsoid $\cE=\{e\mid e^\top \Phi e\leq 1\}$, with $\Phi={\tiny\begin{bmatrix}
     3.4644  &  3.8069\\
    3.8069  &  5.6494
\end{bmatrix} }$, by solving \eqref{eq:optimizaton_thm2}. The problem was formulated in YALMIP and solved using the SDP solver SEDUMI, with a grid search over 210 pairs \((\lambda_0, \lambda_1)\), within 37 seconds. Note that $\cE$ is a PR, $\mathscr{E}_{1-\theta}^{1:N}(e(t))$, by Theorem \ref{thm:Spro}. For $K_{\mathrm{ind}}$, we obtain a PR, $\mathscr{E}_{0.95}^{0:N-1}(Ke(t))=\mathbb{B}(0.4408)$, by Lemma \ref{lemma:prediction_regions}.

We use the PRs, \( \mathscr{E}_{1-\theta}^{1:N}(e(t)) \) and \( \mathscr{E}_{1-\theta}^{0:N-1}(Ke(t)) \), from the direct and indirect methods to tighten the constraints in \eqref{eq:state_constraints}--\eqref{eq:input_constraints} and solve the deterministic problem in \eqref{eq:determ_problem}. The average solve time for the problem in \eqref{eq:determ_problem} on a laptop with an Intel i7-1185G7 processor and 32 GB of RAM is less than 0.02 seconds using the MOSEK solver in YALMIP. Using the resulting solutions, we test constraint satisfaction for $10^4$ new disturbance realizations. For the direct method, state and input constraints are satisfied with frequencies of \( 99.11\% \) and \( 100\% \), meeting the specified probability. For the indirect method, state and input constraints are satisfied for all sampled disturbances. Noisy trajectories are illustrated for both methods in Fig. \ref{fig:PRs}.

The conservatism of the indirect method is expected, as the dashed blue ellipsoid in Fig. \ref{fig:PRs} (right) is a PR for the error state \( e^{(0)}(t) \) for all \( t > 0 \) and any disturbance within \( \mathscr{E}_{0.95}(w(t)) \). In contrast, the direct method synthesizes a gain \( K_{\mathrm{dir}} \) from available training disturbance sequences. 
The merit of our approach is that the two methods can be used independently or in combination. With long horizons, it may be advantageous to collect disturbance samples and solve the problem in \eqref{eq:optimizaton_thm2}, followed by determining PRs using Lemma \ref{lemma:prediction_regions}.

\textbf{Comparison with SO:} We evaluate our method's efficiency against the randomized approach from \cite{Prandini2012}, which uses disturbance feedback parameterization \cite{Goulart2006} to maintain a convex scenario-based relaxation. By enforcing a time-invariant feedback structure (see \cite[Sec. III]{Prandini2012}), we generate scenarios for the constraints in \eqref{eq:dynamics_prob}-\eqref{eq:input_constraints} and solve the scenario optimization with 100 scenarios within 1.5 minutes using the MOSEK solver. However, achieving the specified probabilistic constraints, e.g., with confidence level \(\beta=10^{-3}\) requires at least 5,739 scenarios by \cite[Theorem 1]{Prandini2012} and \cite[Theorem 1]{HewingLCSS2020}, which demands computational time exceeding one hour using our software. In contrast, our synthesis approach in \eqref{eq:Ktrain} or \eqref{eq:optimizaton_thm2} does not impose a minimum number of training scenarios. The desired confidence can be achieved offline through calibration by Lemma \ref{lemma:prediction_regions} for any number of scenarios. Notably, calibration for 5,739 scenarios can be completed within 4 seconds on our computer. The software used in this section is on \cite{Vlahakis_LCSS2024_github}. 

\section{Conclusion}
We have addressed an optimal control problem for linear stochastic systems with unknown distributions using conformal prediction. We propose two methods to design a feedback controller from data that keeps the error state within a prediction region. We provide guarantees independent of data size through a two-step training-calibration process. In future work, we will extend this approach to a stochastic model predictive control setting, exploring closed-loop properties.

\bibliographystyle{IEEEtran}
\bibliography{biblio}

\begin{thebibliography}{10}
\providecommand{\url}[1]{#1}
\csname url@samestyle\endcsname
\providecommand{\newblock}{\relax}
\providecommand{\bibinfo}[2]{#2}
\providecommand{\BIBentrySTDinterwordspacing}{\spaceskip=0pt\relax}
\providecommand{\BIBentryALTinterwordstretchfactor}{4}
\providecommand{\BIBentryALTinterwordspacing}{\spaceskip=\fontdimen2\font plus
\BIBentryALTinterwordstretchfactor\fontdimen3\font minus
  \fontdimen4\font\relax}
\providecommand{\BIBforeignlanguage}[2]{{%
\expandafter\ifx\csname l@#1\endcsname\relax
\typeout{** WARNING: IEEEtran.bst: No hyphenation pattern has been}%
\typeout{** loaded for the language `#1'. Using the pattern for}%
\typeout{** the default language instead.}%
\else
\language=\csname l@#1\endcsname
\fi
#2}}
\providecommand{\BIBdecl}{\relax}
\BIBdecl

\bibitem{Mesbah2016}
A.~Mesbah, ``Stochastic model predictive control: An overview and perspectives
  for future research,'' \emph{IEEE Control Systems Magazine}, vol.~36, no.~6,
  pp. 30--44, 2016.

\bibitem{Kouvaritakis2010explicit}
B.~Kouvaritakis, M.~Cannon, S.~V. Raković, and Q.~Cheng, ``Explicit use of
  probabilistic distributions in linear predictive control,''
  \emph{Automatica}, vol.~46, no.~10, pp. 1719--1724, 2010.

\bibitem{CannonTAC2011}
M.~Cannon, B.~Kouvaritakis, S.~V. Raković, and Q.~Cheng, ``Stochastic tubes in
  model predictive control with probabilistic constraints,'' \emph{IEEE
  Transactions on Automatic Control}, vol.~56, no.~1, pp. 194--200, 2011.

\bibitem{HewingCDC2018}
L.~Hewing and M.~N. Zeilinger, ``Stochastic model predictive control for linear
  systems using probabilistic reachable sets,'' in \emph{2018 IEEE Conf. on
  Decision and Control (CDC)}, 2018, pp. 5182--5188.

\bibitem{KohlerLCSS2022}
J.~Köhler and M.~N. Zeilinger, ``Recursively feasible stochastic predictive
  control using an interpolating initial state constraint,'' \emph{IEEE Control
  Systems Letters}, vol.~6, pp. 2743--2748, 2022.

\bibitem{VlahakisCDC24}
\BIBentryALTinterwordspacing
E.~E. Vlahakis, L.~Lindemann, P.~Sopasakis, and D.~V. Dimarogonas,
  ``Probabilistic tube-based control synthesis of stochastic multi-agent
  systems under signal temporal logic,'' 2024, [accepted to CDC24]. [Online].
  Available: \url{https://arxiv.org/abs/2405.02827}
\BIBentrySTDinterwordspacing

\bibitem{Prandini2012}
M.~Prandini, S.~Garatti, and J.~Lygeros, ``A randomized approach to stochastic
  model predictive control,'' in \emph{2012 IEEE Conf. on Decision and Control
  (CDC)}, 2012, pp. 7315--7320.

\bibitem{LorenzenTAC2017}
M.~Lorenzen, F.~Dabbene, R.~Tempo, and F.~Allgöwer, ``Constraint-tightening
  and stability in stochastic model predictive control,'' \emph{IEEE
  Transactions on Automatic Control}, vol.~62, no.~7, pp. 3165--3177, 2017.

\bibitem{HewingLCSS2020}
L.~Hewing and M.~N. Zeilinger, ``Scenario-based probabilistic reachable sets
  for recursively feasible stochastic model predictive control,'' \emph{IEEE
  Control Systems Letters}, vol.~4, no.~2, pp. 450--455, 2020.

\bibitem{Calafiore2005}
G.~Calafiore and M.~C. Campi, ``{Uncertain convex programs: randomized
  solutions and confidence levels},'' \emph{Mathematical Programming}, vol.
  102, no.~1, pp. 25--46, 2005.

\bibitem{Campi2008}
M.~C. Campi and S.~Garatti, ``The exact feasibility of randomized solutions of
  uncertain convex programs,'' \emph{SIAM Journal on Optimization}, vol.~19,
  no.~3, pp. 1211--1230, 2008.

\bibitem{vovk2005algorithmic}
V.~Vovk, A.~Gammerman, and G.~Shafer, \emph{Algorithmic Learning in a Random
  World}.\hskip 1em plus 0.5em minus 0.4em\relax Springer Science \& Business
  Media, 2005.

\bibitem{shafer2008tutorial}
G.~Shafer and V.~Vovk, ``A tutorial on conformal prediction,'' \emph{Journal of
  Machine Learning Research}, vol.~9, no.~3, pp. 371--421, 2008.

\bibitem{angelopoulos2022gentle}
\BIBentryALTinterwordspacing
A.~N. Angelopoulos and S.~Bates, ``A gentle introduction to conformal
  prediction and distribution-free uncertainty quantification,'' 2022.
  [Online]. Available: \url{https://arxiv.org/abs/2107.07511}
\BIBentrySTDinterwordspacing

\bibitem{zhao2024}
\BIBentryALTinterwordspacing
Y.~Zhao, X.~Yu, J.~V. Deshmukh, and L.~Lindemann, ``Conformal predictive
  programming for chance constrained optimization,'' 2024. [Online]. Available:
  \url{https://arxiv.org/abs/2402.07407}
\BIBentrySTDinterwordspacing

\bibitem{Muthali2023}
A.~Muthali, H.~Shen, S.~Deglurkar, M.~H. Lim, R.~Roelofs, A.~Faust, and
  C.~Tomlin, ``Multi-agent reachability calibration with conformal
  prediction,'' in \emph{62nd IEEE Conf. on Decision and Control (CDC)}, 2023,
  pp. 6596--6603.

\bibitem{Stamouli24}
C.~Stamouli, L.~Lindemann, and G.~Pappas, ``Recursively feasible
  shrinking-horizon {MPC} in dynamic environments with conformal prediction
  guarantees,'' in \emph{Proc. of the 6th Annual Learning for Dynamics \&
  Control Conf.}, vol. 242.\hskip 1em plus 0.5em minus 0.4em\relax PMLR, 2024,
  pp. 1330--1342.

\bibitem{LindemannRAL2023}
L.~Lindemann, M.~Cleaveland, G.~Shim, and G.~J. Pappas, ``Safe planning in
  dynamic environments using conformal prediction,'' \emph{IEEE Robotics and
  Automation Letters}, vol.~8, no.~8, pp. 5116--5123, 2023.

\bibitem{lindemann2024CPsurvey}
\BIBentryALTinterwordspacing
L.~Lindemann, Y.~Zhao, X.~Yu, G.~J. Pappas, and J.~V. Deshmukh, ``Formal
  verification and control with conformal prediction,'' 2024. [Online].
  Available: \url{https://arxiv.org/abs/2409.00536}
\BIBentrySTDinterwordspacing

\bibitem{Goulart2006}
P.~J. Goulart, E.~C. Kerrigan, and J.~M. Maciejowski, ``Optimization over state
  feedback policies for robust control with constraints,'' \emph{Automatica},
  vol.~42, no.~4, pp. 523--533, 2006.

\bibitem{TibshiraniNeurIPS2019}
R.~J. Tibshirani, R.~F. Barber, E.~J. Cand{\`e}s, and A.~Ramdas, ``Conformal
  prediction under covariate shift,'' in \emph{Advances in Neural Information
  Processing Systems}, vol.~32, 2019.

\bibitem{Vovk2012}
V.~Vovk, ``Conditional validity of inductive conformal predictors,'' in
  \emph{Proc. of the Asian Conf. on Machine Learning}, vol.~25, 2012, pp.
  475--490.

\bibitem{boyd2004convex}
S.~Boyd and L.~Vandenberghe, \emph{Convex optimization}.\hskip 1em plus 0.5em
  minus 0.4em\relax Cambridge university press, 2004.

\bibitem{Blanchini1999invariance}
F.~Blanchini, ``Set invariance in control,'' \emph{Automatica}, vol.~35,
  no.~11, pp. 1747--1767, 1999.

\bibitem{Polik2007}
I.~P\'{o}lik and T.~Terlaky, ``A survey of the {S}-lemma,'' \emph{SIAM Review},
  vol.~49, no.~3, pp. 371--418, 2007.

\bibitem{Vlahakis_LCSS2024_github}
E.~E. Vlahakis, \url{https://github.com/lefterisvl83/Vlahakis_LCSS2024}.

\end{thebibliography}

\end{document}